\documentclass[aps,prl,twocolumn,groupedaddress,showpacs]{revtex4}
\usepackage{graphicx}

\begin{document}


\title{Relaxing in foam}


\author{A.D. Gopal and D.J. Durian}
\affiliation{UCLA Department of Physics \& Astronomy, Los Angeles, CA 90095-1547}


\date{\today}

\begin{abstract}
	We investigate the linear mechanical response of an aqueous
	foam, and its relation to the microscopic rearrangement
	dynamics of the bubble-packing structure.  At rest, even
	though the foam is coarsening, the rheology is demonstrated to
	be linear.  Under flow, shear-induced rearrangements compete
	with coarsening-induced rearrangements.  The macroscopic
	consequences are captured by a novel rheological method in
	which a step-strain is superposed on an otherwise steady flow.
\end{abstract}

\pacs{82.70.Rr, 83.60.Rs, 83.80.Iz, 83.85.Cg}

\maketitle



Aqueous foams are tightly packed collections of gas bubbles separated
by a continuous liquid phase \cite{prudkhan,weairehut}.  Like elastic
solids, bulk foams resist shear, completely unlike the gases and
liquids from which the they are comprised.  The origin of this
striking behavior is that the bubbles are jammed, unable to flow
around one another and explore configuration space under thermal
energy.  Thus bubbles distort, rather than rearrange, when subjected
to small shear deformations.  The resulting extra internal gas-liquid
surface area costs energy in proportion to the surface tension, and
this provides a restoring force.  The shear modulus is roughly surface
tension divided by bubble radius, depending precisely on the volume
fraction, $\varepsilon$, of the continuous liquid phase
\cite{princen,mason95b,arnaud99}.  As a foam is made wetter, the
bubbles become progressively rounder, and the shear modulus decreases. 
The elasticity completely vanishes at the point where the bubbles are
close-packed spheres.  This is one example of
unjamming~\cite{liunagel01}.

In this paper we explore other ways to unjam the bubbles in a foam. 
Each could correspond to a different trajectory in a global jamming
phase diagram \cite{liunagel98}.  For example, one could imagine
raising the temperature so that $k_{B}T$ is greater than $\sigma
R^{2}$, which would allow the bubbles to rearrange thermally like
Brownian particles.  However, a typical value is $\sigma
R^{2}/k_{B}\approx 10^{12}$~K; therefore, foams are athermal
far-from-equilibrium systems and raising the temperature is not
feasible.  But there are at least two other ways to drive the system
so that bubble rearrangements occur.  One is through coarsening, the
diffusion of gas from smaller to larger bubbles.  As this proceeds,
local stress inhomogeneities repeatedly build up to some theshold and
relax by sudden avalanche-like local rearrangements.  Such microscopic
dynamics have been observed previously via diffusing-wave spectroscopy
(DWS) \cite{djd91,gopal97,cohen01}.  This raises a string of
interesting questions.  How do localized coarsening-induced
rearrangements affect the linear shear rheology of bulk foams?  Is the
effect similar to thermally excited dynamic heterogeneities in glassy
systems?  An entirely different way to induce rearrangments is by
shear.  Here local stresses also accrue and relax, but in a more
correlated manner.  These rearrangment dynamics have similarly been
studied by DWS \cite{earnshaw94,gopal95,gopal99}.  How does the
elastic character of foam vanish as the shear rate is increased?  Can
we think of the two driving mechanisms as ``heating'' and ultimately
unjamming the sample?  To make progress, we study the linear
mechanical response of an aqueous foam at very long times and low
frequencies.  In a novel twist, we also study linear response {\it
during} uniform shear flow.

Our samples are a commercial aqueous foam, consisting of
nearly-spherical polydisperse gas bubbles, 92\% by volume, that are
tightly packed in an aqueous solution of stearic acid and
triethanolamine (Gillette Company, Boston MA).  Samples are surrounded
by a water bath held at $25.0^{\circ}$C and are measured after the
foam has aged for approximately 100 minutes.  By this time, the
average bubble size is approximately 60 microns and is growing
reproducibly via coarsening \cite{djd91b}; stresses due to the loading
process have also relaxed.  Test durations are sufficiently short that
gravitational liquid drainage and bubble coalescence are negligible.

Our measurements are performed with a Paar~Physica UDS~200 rheometer,
controlling the rotational speed and angular displacement of a solid
cylinder whose axis is vertical and concentric with a fixed
surrounding cup.  The sample cell has an inner radius of 20.0~mm, and
a 4.1~mm gap; these dimensions ensure that the foam can be treated as
a bulk material with uniform stress.  To minimize end-flow effects,
the 98~mm long inner cylinder is much shorter than the depth of the
surrounding cup yet much longer than the gap width.  Wall slip is
precluded by coating both the cylinder and cup with a fine-grade
sandpaper.  Samples are loaded through a 5~mm hole at the bottom of
the cup, after lowering the cylinder to the test position.  DWS
measurements indicate the absence of shear-banding and other secondary
flows \cite{gopal99}.  To re-confirm, we compared with a similarly
coated cone-and-plate cell, with a 10~cm radius and $10^{\circ}$ cone
angle.  Identical rheology results were obtained for both cell
configurations, implying that the imposed shear deformation is
uniform.  Diffuse light transmission indicates that bubbles do not
burst, and voids do not form, even under high
shear~\cite{gopal95,gopal99}.

\begin{figure}
\includegraphics[width=3.00in]{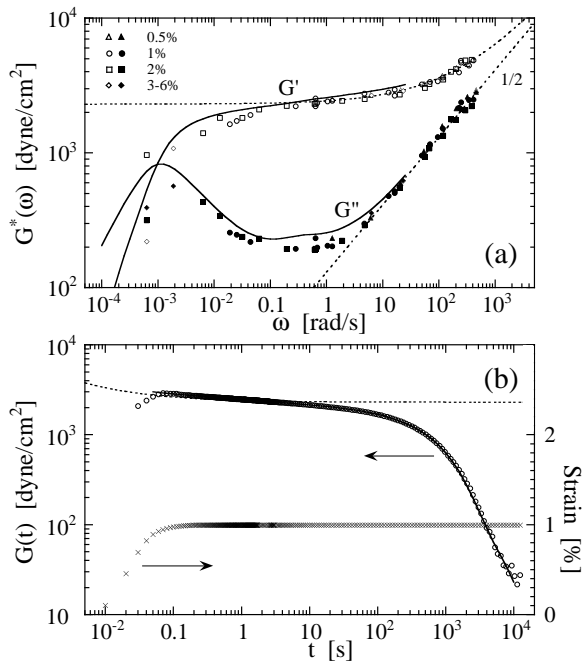}
\caption{Dynamic shear moduli of a coarsening foam, in both the (a)
frequency and (b) time domains; symbols in (a) denote the imposed
strain amplitude.  The dashed curves in (a-b) are a fit to
$G^{*}(\omega)=G_{o}(1+\sqrt{i\omega/\omega_{n}})$ and
$G(t)=G_{\circ}(1+1/\sqrt{\pi \omega_{n}t})$, predicted for nonaffine
bubble motion in the absence of any time evolution.  The solid curve
in (b) is an empirical fit to $G(t)$; appropriately transformed, it
gives the solid curves mathing the storage and loss moduli in (a).}
\label{Glin}
\end{figure}

{\it Effect of coarsening on mechanical response.} To quantify linear
response, we measure both the complex shear modulus,
$G^{*}(\omega)=G'(\omega)+iG''(\omega)$ \cite{ferry,macosko}, and the
stress relaxation modulus, $G(t)$ \cite{ferry,macosko}.  Our data are
displayed in Fig.~\ref{Glin}.  Note that $G'(\omega)>G''(\omega)$,
meaning that the response is primarily elastic rather than
dissipative.  The frequency range for our $G^{*}(\omega)$ data spans
almost six decades, from $2\pi$ divided by sample age (an absolute
minimum below which measurement is not possible) up to a maximum set
by limitations of the rheometer.  Note that different amplitude
strains give the same result, demonstrating absence of wall-slip and
other geometry-dependent artifacts and hinting at linearity of
response.  A more stringent test of linearity is comparison with
$G(t)$, which should be related to $G^{*}(\omega)$ by Fourier
transform \cite{ferry,macosko}.  The time range for our $G(t)$ data
spans over five decades, from the time needed to achieve the step
strain up to the time beyond which stress is zero to within
instrumental limits.  An empirical fit to the $G(t)$ data is shown in
Fig.~\ref{Glin}(b) by a solid curve; this fit is transformed and
plotted in Fig.~\ref{Glin}(a) over a frequency range corresponding to
the time range of the fit.  The agreement is very good, demonstrating
conclusively that the sample is linear.  This is further supported by
an empirical fit to $G^{*}(\omega)$ data at high-$\omega$ and the
comparison of its transform with $G(t)$ data at short-$t$.

Let us now consider the frequency and time dependence of the moduli in
Fig.~\ref{Glin}.  The fit for $\omega>5$~rad/s is to the form
$G^{*}(\omega)=G_{\circ}(1+\sqrt{i\omega/\omega_{n}})$ with free
parameters $G_{\circ}=2300$~dyne/cm$^{2}$ and $\omega_{n}=156$~rad/s. 
The former represents the static shear modulus, roughly surface
tension divided by bubble size~\cite{princen,mason95b,arnaud99}.  The
latter represents the effect of nonaffine deformation of the bubbles
under shear due to local packing configurations that are strong or
weak with respect to the shear direction~\cite{liu96}.  This fit,
including the parameters, is consistent with the $G^{*}(\omega)$ data
in Ref.~\cite{cohen98}, where the frequency range ($0.3-20$~rad/s) was
too small to fully demonstrate the functional form.  According to the
theory of \cite{liu96}, the characteristic frequency is
$\omega_{n}\propto G_{\circ}/\eta_{\infty}$, where the very-high
frequency response is $G^{*}(\omega)=i\eta_{\infty}\omega$.  The
numerical prefactor was not predicted; experimentally, it was found to
depend on $\varepsilon$ and was not of order 1.  To compare, the value
is $\omega_{n}\approx 600$~rad/s for a $\varepsilon=0.38$ emulsion of
0.5~$\mu$m oil droplets in water \cite{mason95a,liu96} (NB: by our
definition, $\omega_{n}$ may be easily read off a plot by locating
where $G'(2\omega_{n})=2G_{\circ}$ and $G''(2\omega_{n})=G_{\circ}$).

The fit to $G_{\circ}(1+\sqrt{i\omega/\omega_{n}})$ fails for
$\omega<5$~rad/s; the corresponding transform fails for $t>20$~s.  At
longer times, the $G(t)$ data decay slowly below $G_{\circ}$, almost
logarithmically, over a few decades before relaxing more
rapidly at around 1000~s.  The transform of this final decay
corresponds to the peak in $G''(\omega)$ at $10^{-3}$~rad/s.  At lower
frequencies, $G^{*}(\omega)$ is unmeasurable; but since the integral
of $G(t)$ over all time is finite, the very-low frequency behavior is
formally $G''(\omega)\propto\omega$ as required by
causality~\cite{buzza}.  Thus the full frequency-dependence of
$G^{*}(\omega)$ for our foam is truly known and well-behaved.  This
resolves a long-standing puzzle \cite{buzza} raised by earlier
measurements \cite{khan,mason95a,cohen98} where $G^{*}(\omega)$ was
roughly constant down to the lowest measured frequencies.

All that remains is to understand the origin of the low-$\omega$ /
long-$t$ behavior.  We contend that evolution of the foam structure by
coarsening is responsible.  One clue is that the onset of deviation
from the high-$\omega$ fits corresponds to the time $\tau_{oq}=20$~s
given by DWS for the time between coarsening-induced rearrangements at
each site.  Another clue is that the final decay of $G(t)$ and,
equivalently, the peak in $G''(\omega)$ correspond to the sample age. 
Since coarsening gives power-law growth, it takes of order the sample
age for the structure to completely change.  It is interesting that
coarsening-induced rearrangments relax microscopic coarsening-induced
stress inhomogeneities far more quickly than the relaxation of
macroscopically-imposed stress.  Rather, the cumulative effect of many
rearrangements and a change in bubble size is needed to relax global
stress.  The net result is a rheology that obeys linear response. 
This is remarkable given that the microscopic relaxation mechanism
isn't thermal motion, but rather evolution.  In effect, coarsening
unjams the foam, so that at low frequencies the rheology is
$G'(\omega) \ll G''(\omega)=\eta\omega$ like an ordinary equilibrium
liquid.

\begin{figure}
\includegraphics[width=3.00in]{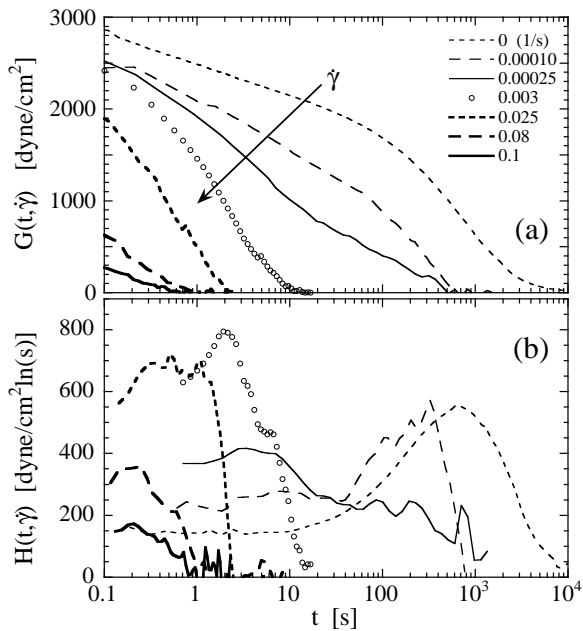}
\caption{(a) Stress relaxation moduli for foams sheared at various 
rates, as listed.  This is given from the stress $\sigma(t)$ following 
the superposition of a step-strain $\Delta\gamma$ on steady shear: 
$G(t,\dot{\gamma})=[\sigma(t)-\sigma(0)]/\Delta\gamma$.  (b) The 
corresponding relaxation spectra, 
$H(t,\dot{\gamma}) \approx dG(t,\dot{\gamma})/d\ln t$.}
\label{Gjog}
\end{figure}

{\it Effect of shear on mechanical response.} Next we investigate how
the application of shear causes a similar loss of elasticity.  For
this we superimpose a small-amplitude step-strain, $\Delta\gamma$, on
top of otherwise steady shear at rate $\dot\gamma$.  The resulting
temporary increase in stress defines the transient stress relaxation
modulus, $G(t,\dot\gamma)=[\sigma(t)-\sigma(0)]/\Delta\gamma$.  A
frequency-domain version of this technique was developed for
polymers \cite{booij}.  Example data are shown in Fig.~\ref{Gjog}(a). 
Note that as the strain rate increases, the elastic character of
the foam melts away.  There are two signs of this.  First, the
progressively smaller intercept, $G(0,\dot\gamma)$, means less elastic
energy storage.  Second, the progressively shorter decay time means
more liquid-like dissipation.  At high enough strain rates, where
there is no transient storage and only dissipation, the foam behaves
like an equilibrium liquid.  Similar behavior has now been observed in
simulations of a model foam \cite{ikono}.

To further quantify the unjamming behavior apparent in
Fig.~\ref{Gjog}(a), we first deduce the relaxation spectrum,
$H(t,\dot{\gamma}) \approx dG(t,\dot{\gamma})/d\ln
t$~\cite{ferry,macosko}.  Results are shown in Fig.~\ref{Gjog}(b). 
For low and zero strain rates, there are two competing relaxation
processes, reflected by a broad peak at very late times and long tail
of short-time modes of nearly equal weight.  The former reflects the
coarsening process, and the latter the nonaffine bubble motion.  As
the shear rate increases, the coarsening peak gradually falls and
another peak gradually rises over the nonaffine tail at short times. 
Presumably this is due to shear-induced rearrangements.

\begin{figure}
\includegraphics[width=3.00in]{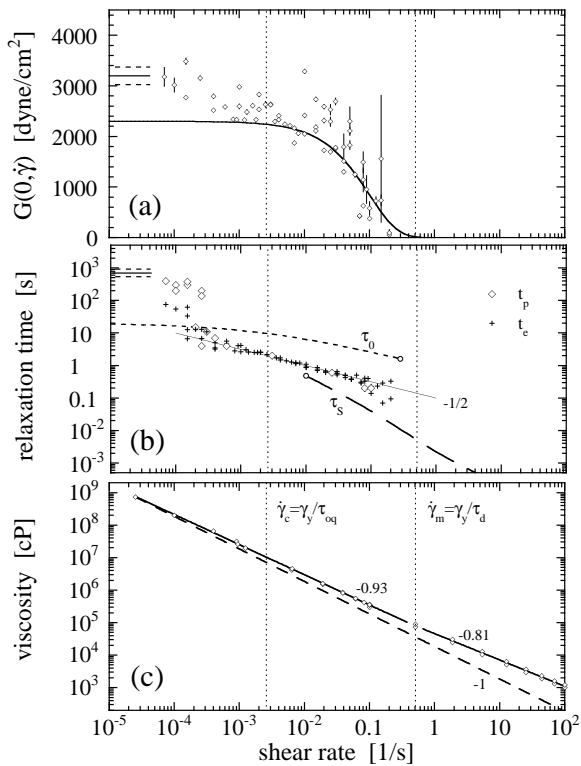}
\caption{(a) Transient shear modulus, (b) transient shear relaxation
times, and (c) viscosity, all as a function of shear rate.  The
vertical dotted lines denote the characteristic shear rates set by the
yield strain divided respectively by the time between rearrangements
and the duration of rearrangements in a quiescent foam.  In (a), the
the solid curve is $G_{0}\exp[-\dot{\gamma}/(0.1{\rm s}^{-1})]$ where
$G_{0}=2300$~dyne/cm$^{2}$ is the shear modulus.  In (b), the open
diamonds indicate where $H(t,\dot{\gamma})$ is a global maximum and
plusses indicate where $G(t,\dot{\gamma})$ falls to $1/e$ of its
initial value; the solid line is a power law with exponent of $-1/2$;
the dashed curves indicate DWS time scales for rearrangements
($\tau_{o}$) and shear ($\tau_{s}$).}
\label{melt}
\end{figure}

The salient features of the stress relaxation are shown in
Fig.~\ref{melt} vs strain rate.  The first plot is of elastic storage,
$G(0,\dot\gamma)$, the value when the superimposed step-strain is
achieved (below about 0.1s).  This decreases very slowly, and is
nearly constant, for strain rates less than about 0.05/s; for higher
strain rates it abruptly vanishes.  The second plot is of stress
relaxation times.  One such measure is $t_{e}$, when stress falls to
$1/e$ of the initial value.  Another measure is $t_{p}$, where
$H(t,\dot{\gamma})$ reaches a global maximum.  At zero and very low
strain rates, these times are different since there are two competing
relaxation mechanisms (evolution and shear).  For strain rates higher
than about $3\times 10^{-4}$/s, the stress relaxation is essentially
exponential and the two relaxation times are hence indistinguishable. 
In this regime, shear completely dominates the relaxation.  It is
puzzling that the relaxation time decreases with increasing strain
rate as $\dot\gamma^{-1/2}$, since on dimensional grounds one would
have expected $\dot\gamma^{-1}$.

Now we may compare the macroscopic rheology with the nature of the
microscopic bubble dynamics.  Previously we used DWS to measure the
strain rate dependence of two microscopic time scales: $\tau_{o}$, the
time between localized discrete rearrangements, and $\tau_{s}$, the
time for adjacent scattering sites to convect apart by one wavelength
of light \cite{gopal99}.  The observed DWS data are reproduced by the
dashed curves in Fig.~\ref{melt}(b).  For very low strain rates, below
about $3\times10^{-4}$/s, the rearrangements are discrete and the time
between events, $\tau_{o}$, is not noticeably different from the
quiescent value.  The stress relaxation time is much much longer than
$\tau_{o}$ - indicating that the cooperative effect of many
coarsening-induced rearrangements is required.  For slightly higher
rates, the time between events decreases, roughly as
$\dot\gamma^{-1/2}$, just like the stress relaxation time.  Since the
stress relaxation time is now shorter than $\tau_{o}$, not every site
in the foam must rearrange in order to relax the overall stress. 
Shear-induced events begin to dominate over coarsening-induced events
at strain rates near $\dot\gamma_{c}=\gamma_{y}/\tau_{oq}=0.025$/s,
where $\gamma_{y}=0.05$ is the yield strain \cite{gopal99}.  At still
higher rates, events merge together and the flow becomes progressively
more homogeneous and smooth.  The crossover strain rate is
$\dot\gamma_{m}=\gamma_{y}/\tau_{d}=0.5$/s, where $\tau_{d}=0.1$~s is
the duration of rearrangements \cite{gopal99}.  Many physical
quantities display a change in character above and below
$\dot\gamma_{m}$ \cite{onoPRE}.  Indeed, just below $\dot\gamma_{m}$,
the new rheological measures of elasticity vanish at precisely where
we no longer can detect dicrete rearrangments (i.e. at the endpoint of
the dashed curve for $\tau_{o}$).  At higher strain rates, the bubble
motion is dominated entirely by uniform shear, as characterized by the
DWS time scale $\tau_{s}$.  The onset of this correlated shearing
motion of bubbles (i.e. the starting point of the $\tau_{s}$ curve)
corresponds quite well to where $G(0,\dot\gamma)$ begins to drop.

Finally, we note that the dramatic changes in the elastic character of
the foam with strain rate are not reflected very strongly in the
viscosity of the foam.  As seen in Fig.~\ref{melt}(c), the
viscosity decreases across the whole strain rate range, though not
quite as fast as $\dot\gamma^{-1}$ (which would have indicated
constant stress).  There is at most a slight change in exponent at
$\dot\gamma_{m}$.  Similar behavior is found in simulations
\cite{langerEL}.  Altogether this emphasizes the importance of the
superposition method to measure $G(t,\dot\gamma)$, since it provides a
clear dramatic signature of the unjamming transition.  With this tool,
and by comparison with DWS data, we have succeeded in connecting the
nature of microscopic bubble dynamics with the resulting macroscopic
rheological behavior.  We have thereby shown that the unjamming of
foam can be accomplished both by time and by application of shear. 
The unjammed liquid-like state is very similar to what would be
achieved by raising the temperature for a thermal system.  An
important next step would be to deduce an effective temperature,
recently shown in simulation of a sheared model foam to have many of
the attributes of a true statistical mechanical temperature
\cite{onoPRL}.

We thank A.J. Liu and P.T. Mather for helpful discussions.  This 
work was supported by NASA Microgravity Fluid Physics grant NAG3-2481.

\bibliography{JogRefs}

\end{document}